\documentclass[
aps,%
12pt,%
final,%
notitlepage,%
oneside,%
onecolumn,%
nobibnotes,%
nofootinbib,%
superscriptaddress,%
noshowpacs,%
centertags]%
{revtex4}
\usepackage{graphicx}
\usepackage{amssymb}
\usepackage{textcomp}  
\usepackage{caption}
\usepackage{subcaption}
\begin{document}

\title{Neutrino astronomy at Lake Baikal}


\author{V.~A.~Allakhverdyan}
\affiliation{Joint Institute for Nuclear Research, Dubna, Russia}
\author{A.~D.~Avrorin}
\affiliation{Institute for Nuclear Research, Russian Academy of Sciences, Moscow, Russia}
\author{A.~V.~Avrorin}
\affiliation{Institute for Nuclear Research, Russian Academy of Sciences, Moscow, Russia}
\author{V.~M.~Aynutdinov}
\affiliation{Institute for Nuclear Research, Russian Academy of Sciences, Moscow, Russia}
\author{Z.~Barda\v{c}ov\'{a}}
\affiliation{Comenius University, Bratislava, Slovakia}
\affiliation{Czech Technical University in Prague, Prague, Czech Republic}
\author{I.~A.~Belolaptikov}
\affiliation{Joint Institute for Nuclear Research, Dubna, Russia}
\author{E.~A.~Bondarev}
\affiliation{Institute for Nuclear Research, Russian Academy of Sciences, Moscow, Russia}
\author{I.~V.~Borina}
\affiliation{Joint Institute for Nuclear Research, Dubna, Russia}
\author{N.~M.~Budnev}
\affiliation{Irkutsk State University, Irkutsk, Russia}
\author{V.~A.~Chadymov}
\affiliation{Independent researchers}
\author{A.~S.~Chepurnov}
\affiliation{Skobeltsyn Institute of Nuclear Physics MSU, Moscow, Russia}
\author{V.~Y.~Dik}
\affiliation{Joint Institute for Nuclear Research, Dubna, Russia}
\affiliation{Institute of Nuclear Physics of the Ministry of Energy of the Republic of Kazakhstan, Kazakhstan}
\author{G.~V.~Domogatsky}
\affiliation{Institute for Nuclear Research, Russian Academy of Sciences, Moscow, Russia}
\author{A.~A.~Doroshenko}
\affiliation{Institute for Nuclear Research, Russian Academy of Sciences, Moscow, Russia}
\author{R.~Dvornick\'{y}}
\affiliation{Institute for Nuclear Research, Russian Academy of Sciences, Moscow, Russia}
\author{A.~N.~Dyachok}
\affiliation{Irkutsk State University, Irkutsk, Russia}
\author{Zh.-A.~M.~Dzhilkibaev}
\affiliation{Institute for Nuclear Research, Russian Academy of Sciences, Moscow, Russia}
\author{E.~Eckerov\'{a}}
\affiliation{Comenius University, Bratislava, Slovakia}
\affiliation{Czech Technical University in Prague, Prague, Czech Republic}
\author{T.~V.~Elzhov}
\affiliation{Joint Institute for Nuclear Research, Dubna, Russia}
\author{V.~N.~Fomin}
\affiliation{Independent researchers}
\author{A.~R.~Gafarov}
\affiliation{Irkutsk State University, Irkutsk, Russia}
\author{K.~V.~Golubkov}
\affiliation{Institute for Nuclear Research, Russian Academy of Sciences, Moscow, Russia}
\author{N.~S.~Gorshkov}
\affiliation{Joint Institute for Nuclear Research, Dubna, Russia}
\author{T.~I.~Gress}
\affiliation{Irkutsk State University, Irkutsk, Russia}
\author{K.~G.~Kebkal}
\affiliation{LATENA, St. Petersburg, Russia}
\author{V.~K.~Kebkal}
\affiliation{LATENA, St. Petersburg, Russia}
\author{I.~Kharuk}
\affiliation{Institute for Nuclear Research, Russian Academy of Sciences, Moscow, Russia}
\author{E.~V.~Khramov}
\affiliation{Joint Institute for Nuclear Research, Dubna, Russia}
\author{M.~I.~Kleimenov}
\affiliation{Institute for Nuclear Research, Russian Academy of Sciences, Moscow, Russia}
\author{M.~M.~Kolbin}
\affiliation{Joint Institute for Nuclear Research, Dubna, Russia}
\author{S.~O.~Koligaev}
\affiliation{INFRAD, Dubna, Russia}
\author{K.~V.~Konischev}
\affiliation{Institute for Nuclear Research, Russian Academy of Sciences, Moscow, Russia}
\author{A.~V.~Korobchenko}
\affiliation{Joint Institute for Nuclear Research, Dubna, Russia}
\author{A.~P.~Koshechkin}
\affiliation{Institute for Nuclear Research, Russian Academy of Sciences, Moscow, Russia}
\author{V.~A.~Kozhin}
\affiliation{Skobeltsyn Institute of Nuclear Physics MSU, Moscow, Russia}
\author{M.~V.~Kruglov}
\affiliation{Joint Institute for Nuclear Research, Dubna, Russia}
\author{V.~F.~Kulepov}
\affiliation{Nizhny Novgorod State Technical University, Nizhny Novgorod, Russia}
\author{A.~A.~Kulikov}
\affiliation{Irkutsk State University, Irkutsk, Russia}
\author{Y.~E.~Lemeshev}
\affiliation{Irkutsk State University, Irkutsk, Russia}
\author{R.~R.~Mirgazov}
\affiliation{Irkutsk State University, Irkutsk, Russia}
\author{D.~V.~Naumov}
\affiliation{Joint Institute for Nuclear Research, Dubna, Russia}
\author{A.~S.~Nikolaev}
\affiliation{Skobeltsyn Institute of Nuclear Physics MSU, Moscow, Russia}
\author{I.~A.~Perevalova}
\affiliation{Irkutsk State University, Irkutsk, Russia}
\author{D.~P.~Petukhov}
\affiliation{Institute for Nuclear Research, Russian Academy of Sciences, Moscow, Russia}
\author{E.~N.~Pliskovsky}
\affiliation{Joint Institute for Nuclear Research, Dubna, Russia}
\author{M.~I.~Rozanov}
\affiliation{St. Petersburg State Marine Technical University, St. Petersburg, Russia}
\author{E.~V.~Ryabov}
\affiliation{Irkutsk State University, Irkutsk, Russia}
\author{G.~B.~Safronov}
\affiliation{Institute for Nuclear Research, Russian Academy of Sciences, Moscow, Russia}
\author{B.~A.~Shaybonov}
\affiliation{Joint Institute for Nuclear Research, Dubna, Russia}
\author{V.~Y.~Shishkin}
\affiliation{Skobeltsyn Institute of Nuclear Physics MSU, Moscow, Russia}
\author{E.~V.~Shirokov}
\affiliation{Skobeltsyn Institute of Nuclear Physics MSU, Moscow, Russia}
\author{F.~\v{S}imkovic}
\affiliation{Comenius University, Bratislava, Slovakia}
\affiliation{Czech Technical University in Prague, Prague, Czech Republic}
\author{A.~E.~Sirenko}
\affiliation{Joint Institute for Nuclear Research, Dubna, Russia}
\author{A.~V.~Skurikhin}
\affiliation{Skobeltsyn Institute of Nuclear Physics MSU, Moscow, Russia}
\author{A.~G.~Solovjev}
\affiliation{Joint Institute for Nuclear Research, Dubna, Russia}
\author{M.~N.~Sorokovikov}
\affiliation{Joint Institute for Nuclear Research, Dubna, Russia}
\author{I.~\v{S}tekl}
\affiliation{Czech Technical University in Prague, Prague, Czech Republic}
\author{A.~P.~Stromakov}
\affiliation{Institute for Nuclear Research, Russian Academy of Sciences, Moscow, Russia}
\author{O.~V.~Suvorova}
\affiliation{Institute for Nuclear Research, Russian Academy of Sciences, Moscow, Russia}
\author{V.~A.~Tabolenko}
\affiliation{Irkutsk State University, Irkutsk, Russia}
\author{V.~I.~Tretyak}
\affiliation{Joint Institute for Nuclear Research, Dubna, Russia}
\author{B.~B.~Ulzutuev}
\affiliation{Joint Institute for Nuclear Research, Dubna, Russia}
\author{Y.~V.~Yablokova}
\affiliation{Joint Institute for Nuclear Research, Dubna, Russia}
\author{D.~N.~Zaborov}
\thanks{Corresponding author}
\affiliation{Institute for Nuclear Research, Russian Academy of Sciences, Moscow, Russia}
\author{S.~I.~Zavyalov}
\affiliation{Joint Institute for Nuclear Research, Dubna, Russia}
\author{D.~Y.~Zvezdov}
\affiliation{Joint Institute for Nuclear Research, Dubna, Russia}

\begin{abstract}

{\bf Abstract} -- High energy neutrino astronomy has seen significant progress in the past few years. This includes the detection of neutrino flux from the Galactic plane, as well as strong evidence for neutrino emission from the active galaxy NGC 1068, both reported by IceCube. New results start coming from the two km$^3$-scale neutrino telescopes under construction in the Northern hemisphere: KM3NeT in the Mediterranean Sea and Baikal-GVD in Lake Baikal. After briefly reviewing the status of the field, we present the current status of the Baikal-GVD neutrino telescope and its recent results, including observations of atmospheric and astrophysical neutrinos.

\end{abstract}

\maketitle

\section{Introduction}
High energy neutrino astronomy is a branch of astronomy employing neutrinos of GeV--PeV energies as astrophysical messengers.
The use of neutrino provides us with a complementary view of the Universe thanks to its unique properties.
In particular, thanks to the weak interaction of neutrino with matter, neutrino is capable of escaping from dense environments which are opaque to photons.
Neutrino can travel unimpeded through gas and dust and it does not interact significantly with the background optical, infrared or microwave radiation on its way to the observer, unlike high energy photons.
As a result, the cosmos is practically transparent to neutrinos.\footnote{We leave aside the beyond-standard-model scenarios such as, e.g., resonant neutrino self-interactions.}
The neutrino is not affected by magnetic fields\footnote{We ignore here the effects of the non-zero neutrino dipole moment and other electromagnetic properties which are generally predicted to be extremely small.}, and thus it always points to its source, unlike charged cosmic rays.
Finally, it is stable, i.e. it does not decay to other particles on the way to Earth.
These properties make neutrino an excellent astrophysical messenger, complementary to photons, cosmic rays and gravitational waves.
In a typical hadronic emission scenario, high-energy neutrinos are produced in the decays of charged pions and muons, which are in turn produced in $p$-$p$ or $p$-$\gamma$ interactions due to accelerated cosmic rays colliding with matter particles of photons.
Hence detecting neutrino sources allows one to trace the production and acceleration sites of cosmic rays.

In the GeV--TeV energy range the neutrino sky is largely dominated by atmospheric neutrinos -- a side-product of cosmic ray interactions in the Earth's atmosphere.
This atmospheric neutrino background limits the sensitivity of Earth-based neutrino telescopes to astrophysical neutrinos.
With the atmospheric neutrino flux falling steeply with energy, the preferred energy band for observing the astrophysical neutrinos appears to be above $\sim$10~TeV, although observations at lower energies can also be efficient for point-like sources and especially transient sources, thanks to reduced background.
At these energies, typical rates of astrophysical neutrino interactions are just a dozen events per year per cubic kilometer, necessitating the construction of huge ($\gtrsim$1~km$^3$) neutrino telescopes.
Such a large detector volume can be practically achieved using a natural site only.
Particularly suited for that purpose are deep-sea, deep-lake and deep-ice sites characterized by high optical transparency of the natural medium, allowing for the use of the Cherenkov detection technique at large scales \cite{Markov1960}.
Sensitive photo-sensors installed in the natural medium are used to register the Cherenkov light of secondary particles produced by neutrino interactions inside or in the vicinity of the detector.
The amplitudes (light intensities) and relative times of the signals recorded by the three-dimensional array of the photo-sensors are then used to reconstruct the neutrino energy and arrival direction.
The detectors using this technique for high-energy neutrino detection are known as neutrino telescopes.

Typically, neutrino events recorded by neutrino telescopes are classified into two main classes: tracks, which largely correspond to $\nu_\mu$ charged current (CC) interactions; and showers, which correspond to $\nu_e$ CC, $\nu_\tau$ CC, and all-flavor neutral current (NC) interactions.\footnote{This also includes the corresponding anti-neutrino flavours.}
Generally speaking, track-like events offer a better angular resolution (thanks to the kilometer-long muon tracks), while cascade-like events offer a better energy resolution (due to calorimetric measurement of the fully contained particle shower). 
Additionally, a "double-bang" cascade event signature is characteristic of very high energy $\nu_\tau$ CC interactions, where the first cascade is due to the hadronic shower started at the neutrino-nucleon interaction vertex and the second cascade is due to the $\tau$ lepton decay.
Tau neutrinos are rarely produced in the atmosphere, hence any such double-bang $\nu_\tau$ event would likely be of astrophysical origin.

Currently there are three neutrino telescopes in operation or under construction around the world: IceCube \cite{IceCube} at the South Pole, KM3NeT \cite{KM3NeT_LoI} in the Mediterranean Sea, and Baikal-GVD in Lake Baikal.
The smaller-scale ANTARES neutrino telescope \cite{ANTARES}, which operated in the Mediterranean Sea since 2008, was dismantled in February 2022.

The IceCube detector \cite{IceCube} occupies a 1 km$^3$ volume of Antarctic ice at the South Pole. 
It consists of 5160 optical sensors arranged on 86 vertical strings.
The array was completed in 2010. 
In addition to the main array, IceCube has a densely-instrumented core for GeV neutrino studies (DeepCore) and a surface array for air shower detection (IceTop).
IceCube is currently the most sensitive neutrino telescope in the world.
However, its location at the South Pole implies a reduced sensitivity for sources in the Southern celestial hemisphere, due to large background of down-going atmospheric muons (for track-like events).
The angular resolution of IceCube is limited to some extent by significant scattering of light on ice non-uniformities.
IceCube is notable for its discovery of a diffuse astrophysical neutrino flux \cite{IceCube_diffuse} and detailed measurements of that flux in both cascade and track channels (see, e.g., \cite{IceCube_diffuse_2020}).
The origin of this diffuse neutrino flux however remains unknown and is subject to much debate.
IceCube has reported statistically significant excesses for two point-like sources: the blazar TXS~0506+056 \cite{IceCube_TXS0506,IceCube_TXS0506_flare} and the Seyfert galaxy NGC~1068 \cite{IceCube_NGC1068}.
Recently, IceCube also reported an excess of events from the Galactic plane \cite{IceCube_Galactic_diffuse}.

The KM3NeT-ARCA detector \cite{KM3NeT_LoI} is a 1 km$^3$ volume detector under construction in the Mediterranean Sea near Sicily (Italy). When completed, it will consist of 4140 optical sensors installed on 230 strings. KM3NeT uses a highly efficient multi-PMT design for its optical sensors.
As of writing this manuscript, only the first 33 strings have been installed and are operational in ARCA.
The KM3NeT-ORCA detector, under construction near the French Mediterranean coast, is a densely instrumented version of ARCA, optimized for the neutrino mass hierarchy measurement \cite{KM3NeT_LoI}. Due to the relatively small volume ($\sim$0.008 km$^3$), ORCA has a limited sensitivity to TeV--PeV neutrinos.
Thanks to excellent optical properties (low absorption and low scattering) of the deep Mediterranean water, KM3NeT will provide a substantial improvement over the angular resolution of IceCube.

Baikal-GVD is a 1 km$^3$ detector under construction in Lake Baikal (Russia).
It offers a sky coverage which is in part complementary to that of IceCube, along with a better angular resolution and a similarly large effective volume.
The Baikal-GVD detector is described in the Section \ref{section2} and its main results are presented in Section \ref{section3}.
Section \ref{sect:conclusion} concludes the paper.

\section{Baikal-GVD}
\label{section2}
The Baikal Gigaton Volume Detector (Baikal-GVD) is a cubic-kilometer scale underwater neutrino detector currently under construction in Lake Baikal, Russia.
The Baikal-GVD detector site is located in the Southern basin of Lake Baikal at 51$^\circ$\,46'\,N 104$^\circ$\,24'\,E, 3.6 km offshore and 1365~m deep.
The light absorption length in the deep lake water reaches 22~m \cite{Baikal_optical_water_properties}.
The light scattering length is between 30~m and 50~m.
After accounting for the forward-peaked diagram of light scattering, the effective scattering length is up to 480~m.
The lake is covered with thick ice (up to about 1~m) from February to mid-April,
providing a convenient solid platform for massive detector deployment and maintenance operations.
The experiment is aimed to provide observations of the neutrino sky with a sensitivity similar to that of IceCube and a complementary field of view.
Simultaneous operation with KM3NeT and IceCube allows for continuous monitoring of transient phenomena over the full sky and improved all-sky combined sensitivity.

\begin{figure}
  \centering
  \includegraphics[height=10.0cm]{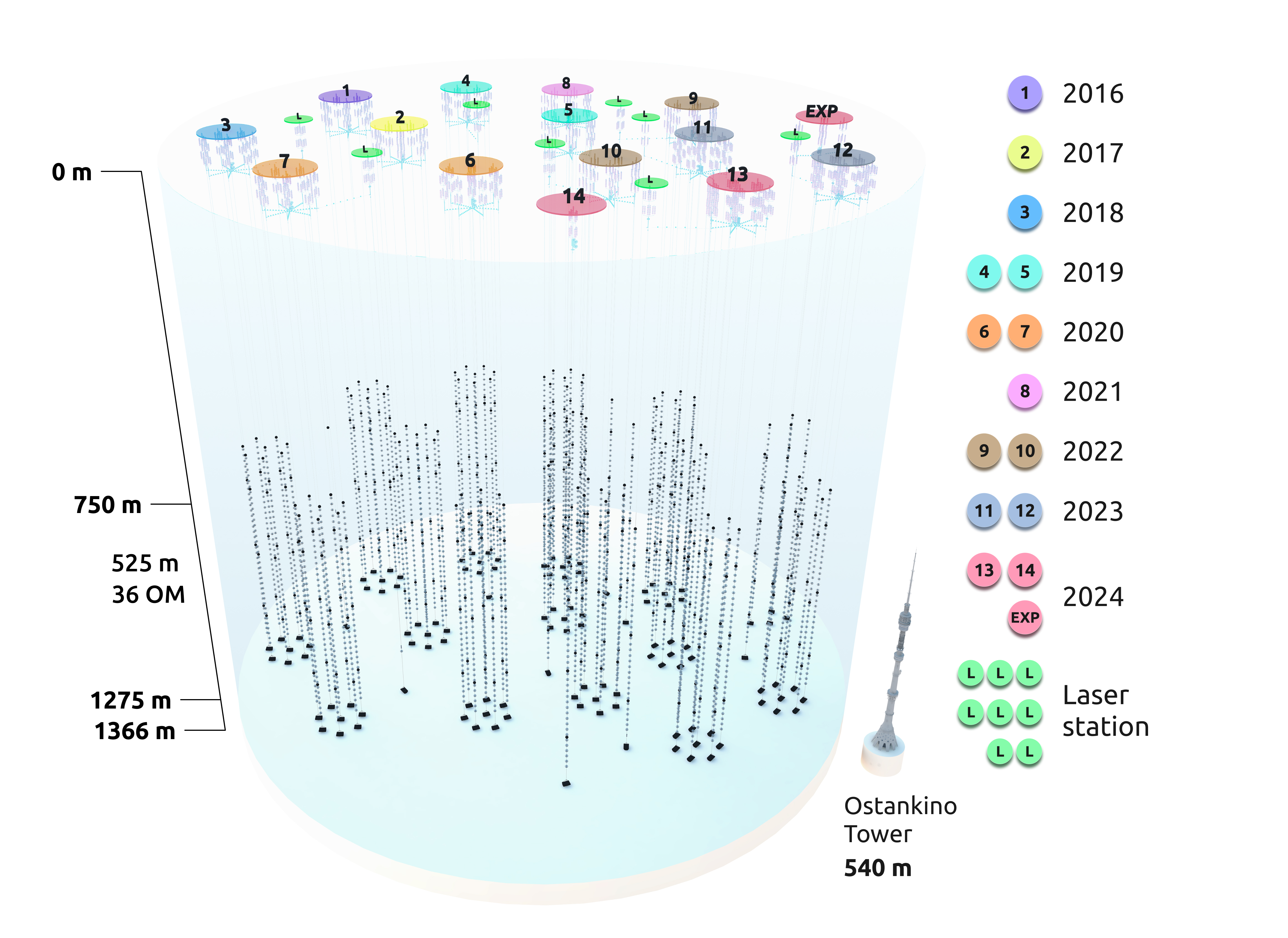}
  \caption{Schematic view of the Baikal-GVD detector (see text).}
  \label{fig:baikal_gvd}
\end{figure}

The Baikal-GVD layout is shown in Fig.~\ref{fig:baikal_gvd}. 
The basic sensitive component of the Baikal-GVD detector is the optical module (OM).
The OM includes a 10-inch high-quantum-efficiency PMT (Hamamatsu R7081-100), a high voltage unit and front-end electronics, all enclosed in a pressure-resistant glass sphere.
The OM is also equipped with calibration LEDs, a tiltmeter/accelerometer and a compass.
The OMs are arranged on vertical strings.
Each string holds 36 optical modules (OMs) spaced every 15 m along the string at depths between 750~m and 1275~m below surface.
The optical modules are electrically connected to special electronics modules, which are, like OMs, housed in glass spheres and attached to the string's load-carrying cable.
Strings are anchored to the lakebed and kept vertical by buoys at their tops, as well as by the buoyancy of the optical and electronics modules.
The strings also hold hydrophones for acoustic positioning and LED beacons for calibration  \cite{Baikal_calibration,Baikal_positioning}.
The Baikal-GVD strings are grouped in clusters, 8 strings per cluster, with 60 m horizontal spacing between the strings.
Additional dedicated strings equipped with high-power pulsed lasers are installed in-between the GVD clusters. These are used for detector calibration and light propagation studies.
Some of these additional strings also carry a full set of 36 OMs, thus acting as "inter-cluster" strings.
All strings of a given cluster are connected to the cluster center module assembly, which is attached near the top of the central string of the cluster.
The cluster center is connected to the Baikal-GVD shore station via a dedicated electro-optical cable.

The effective volume of a stand-alone GVD cluster for cascade-like neutrino events with energy above 100 TeV is $\approx$0.04--0.05 km$^3$.
The effective volume grows almost linearly with the number of clusters, thus a 20-cluster version of Baikal-GVD would have an effective volume approaching 1 km$^3$ at E$=$1 PeV.
This is similar to the effective volume of the 230-string version of KM3NeT-ARCA, although at lower energies KM3NeT-ARCA will generally have a larger efficiency than Baikal-GVD \cite{Zaborov_Kleimenov_2024}.

The first cluster of Baikal-GVD was deployed in 2016.
Since then, one or two new clusters were added every year.
Today Baikal-GVD consists of 13 completed clusters, plus several additional inter-cluster strings, laser station strings, and experimental ("fiber-optic DAQ") strings.
The detector occupies a water volume of $\approx$0.6 km$^3$.
As it stands, Baikal-GVD is currently the largest neutrino telescope in the Northern Hemisphere.
The Baikal-GVD construction plan anticipates the addition of at least three new clusters in the next two years.

\section{Main results from Baikal-GVD}
\label{section3}
Several physics analyses are currently ongoing in Baikal-GVD.
This includes the measurements of atmospheric muons, atmospheric neutrinos and searches for astrophysical neutrino fluxes.

\begin{figure}
  \centering
  \includegraphics[width=8.0cm]{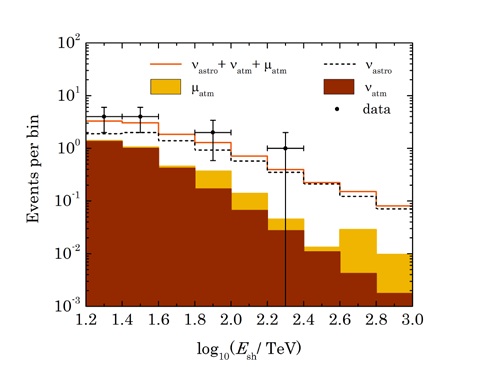}
  \includegraphics[width=8.0cm]{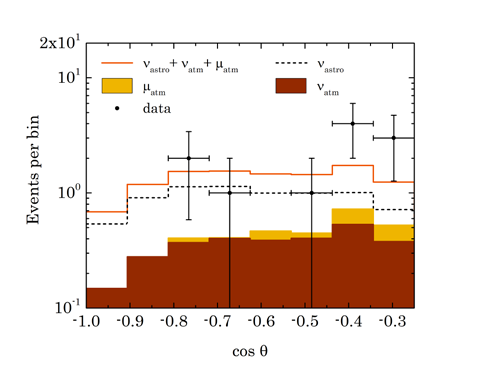}
  \caption{Reconstructed cascade energy (left panel) and zenith angle (right panel) distributions of the 
upward-going cascade-like neutrino events found in the 2018--2021 Baikal-GVD data. Black points are data, with statistical uncertainties. The best-fit
distribution of astrophysical neutrinos (dashed line), expected distributions from atmospheric muons
(yellow) and atmospheric neutrinos (brown) and the sum of the expected signal and background
distributions (orange line) are also shown. The atmospheric background histograms are stacked (filled
colors). Plots from \cite{Baikal_cascades}.}
  \label{fig:cascade_analysis_results_1}
\end{figure}

\begin{figure}
  \centering
  \includegraphics[height=6.5cm]{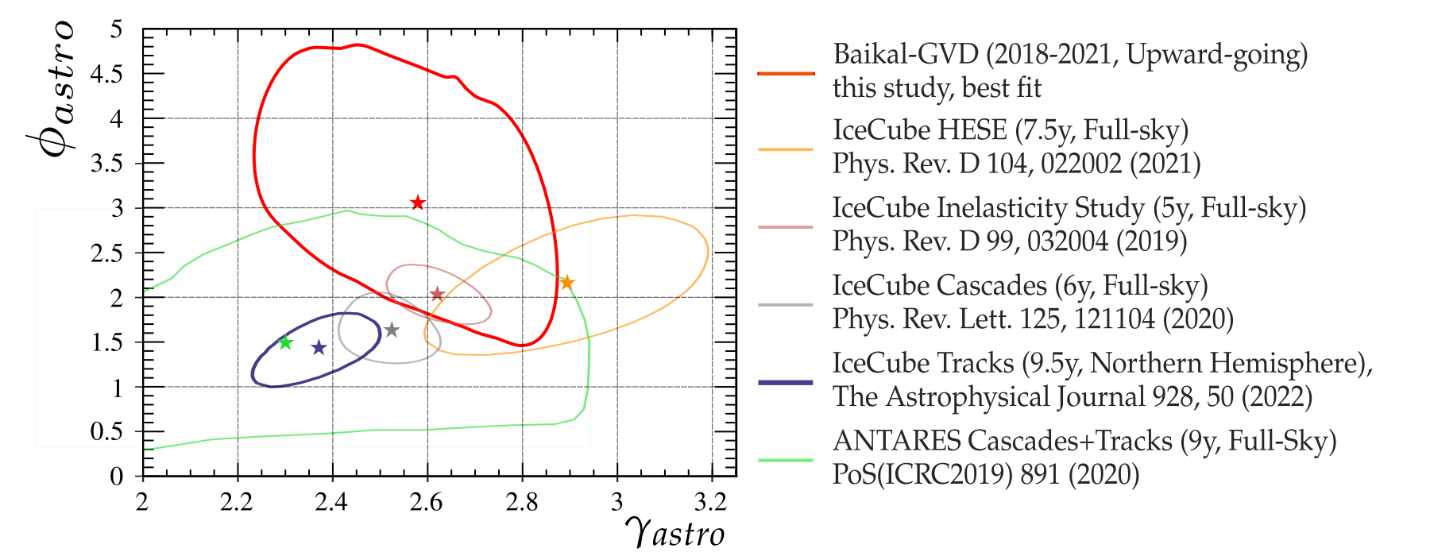}
  \caption{The best fit parameters and the contours of the 68\% confidence region (red curve) for the single
power law hypothesis obtained in the upward-going cascade analysis of the 2018--2021 Baikal-GVD data. Other best
fits are shown for studies based on IceCube high-energy starting events (orange curve, cascade-like events
(gray curve), an inelasticity study (purple curve) and track-like events (blue curve) and ANTARES observation in a combined study of tracks and cascades (green curve). Plot taken from \cite{Baikal_cascades}.}
  \label{fig:cascade_analysis_results_2}
\end{figure}

The first observation of the diffuse cosmic neutrino flux with Baikal-GVD was achieved using a high-energy cascade analysis \cite{Baikal_cascades}.
The analysis is primarily sensitive to neutrinos and anti-neutrinos of electron and tau flavors.
Using cascade-like events collected by Baikal-GVD in 2018--2021, a significant excess of
events over the expected atmospheric background is observed.
The reconstructed energy and zenith angle distributions are shown in Fig.~\ref{fig:cascade_analysis_results_1}.
The null cosmic flux assumption is rejected with a significance of 3.05$\,\sigma$.
Assuming a single power law model of the astrophysical neutrino flux with identical contribution from each neutrino flavor, the following best-fit parameter values are found: the spectral index $\gamma=2.58^{+0.27}_{-0.33}$
and the flux normalization $\Phi_{astro}=3.04^{+1.52}_{-1.21}$ per one flavor at 100 TeV.
The spectral measurement contours are shown in Fig.~\ref{fig:cascade_analysis_results_2}.
The observed flux parameters are consistent with the high-energy diffuse cosmic neutrino flux observations by IceCube.

A rare neutrino event with an estimated energy of 224 $\pm$ 75 TeV was found to come from a direction which is within the error circle from the direction of TXS 0506+056 \cite{Baikal_TXS_cascade}. The event was registered in April 2021 and at that time was the highest energy cascade detected so far by the Baikal-GVD neutrino telescope from a direction below horizon.

\begin{figure}
  \centering
  \includegraphics[trim=0 100 0 150,clip,height=7.5cm]{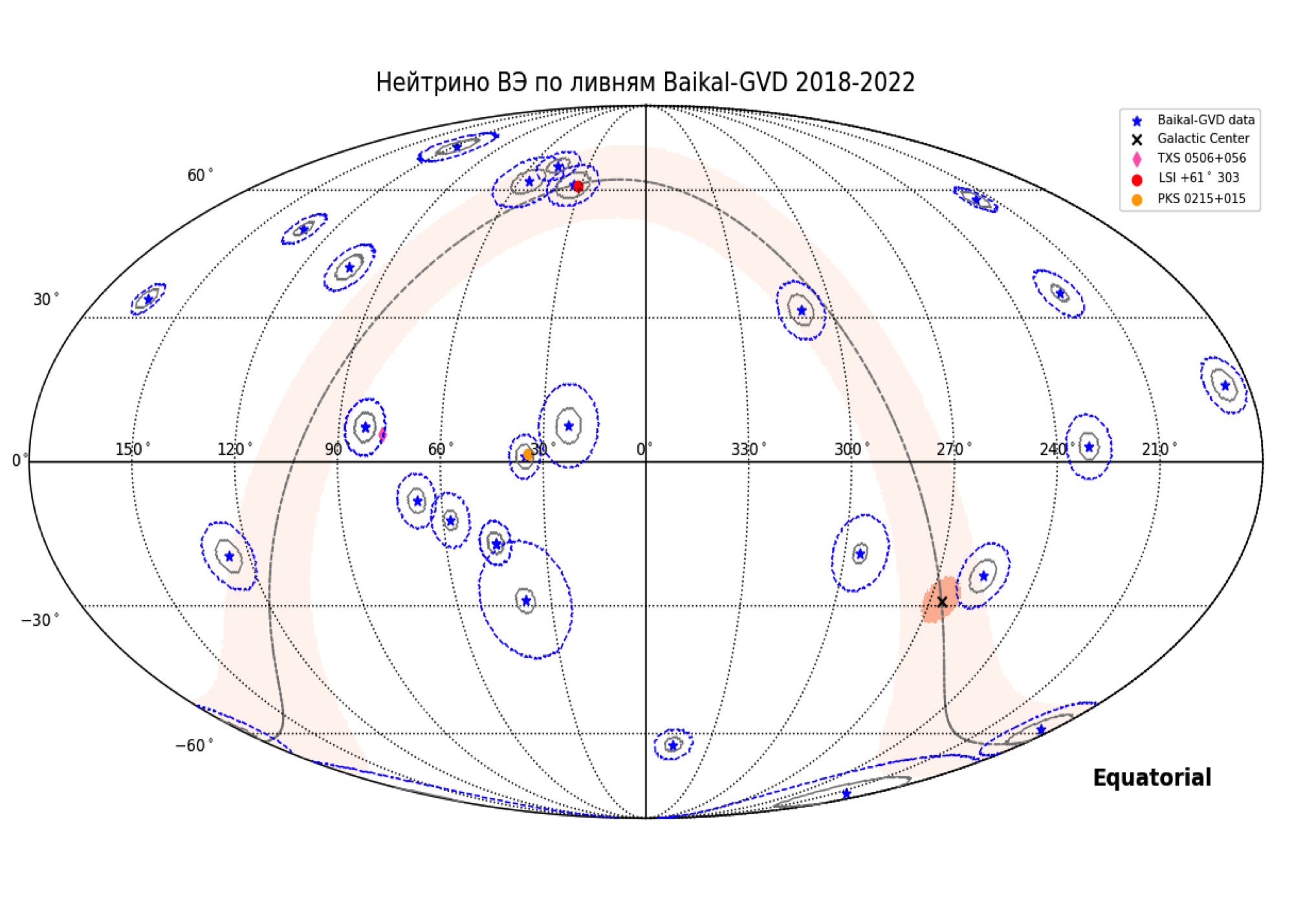}
  \caption{Arrival directions of high-energy and under-horizon Baikal-GVD neutrino cascades (stars) and their uncertainty (50 per cent and 90 per cent) regions (ellipses) in the sky map (equatorial coordinates). Data from 2018--2022. Positions of some notable objects are also shown. The Galactic plane is indicated by the black curve.}
  \label{fig:baikal_cascade_sky}
\end{figure}

Using the Baikal-GVD cascade data set collected in 2018--2022, some additional potential source associations were found, including LS I+61 303 and Swift J0243.6+6124, which both fall within the 90\% uncertainty
region of an event doublet (see Fig.~\ref{fig:baikal_cascade_sky}) \cite{Baikal_cascades_MNRAS_2023}.

\begin{figure}
  \centering
  \includegraphics[width=8.0cm]{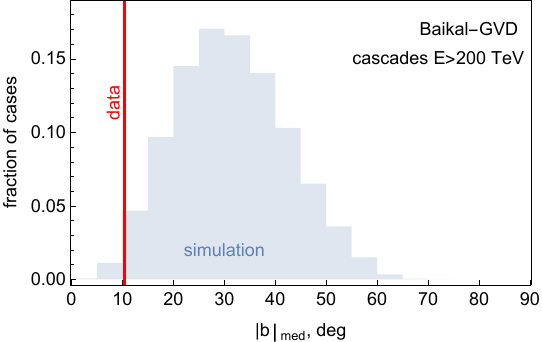}
  \includegraphics[width=7.7cm]{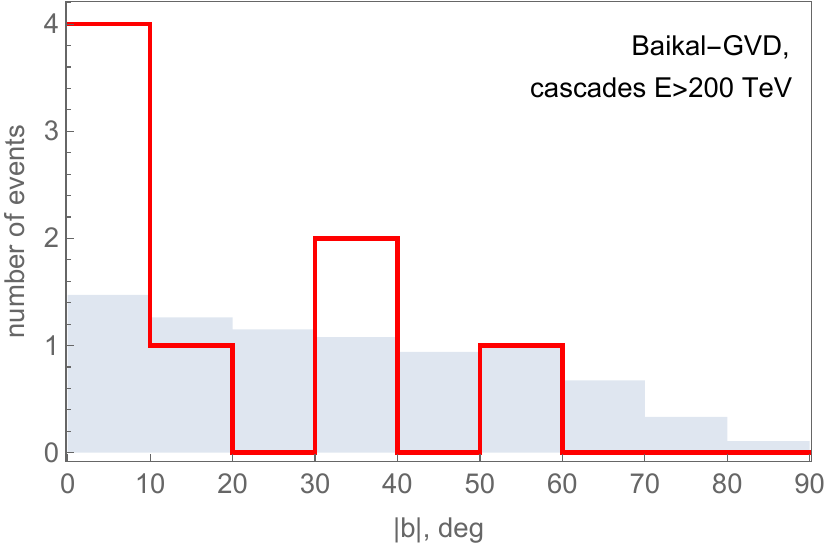}
  \caption{Results of a search for a galactic neutrino excess using Baikal-GVD cascade-like events collected in 2018--2023.
  Left: Distribution (shaded histogram) of the median of the absolute value of galactic latitude $|b|_{med}$ in simulated sets of Baikal-GVD cascades with E$>$200~TeV. The observed value is shown by the vertical red line.
  Right: Observed (red line) and expected (shaded histogram) distribution of $|b|$ for Baikal-GVD cascades with E$>$200~TeV.
  Figures from \cite{Baikal_Galactic_median}.
  }
  \label{fig:baikal_cascade_galactic}
\end{figure}

Using a sample of cascade events with estimated neutrino energies above 200 TeV, detected by the partially deployed Baikal-GVD in 2018--2023, a statistical test for correlation with the galactic plane has been performed. For this, the median of the absolute value of galactic latitude was used as a test statistic. The analysis suggests a 2.5$\,\sigma$ excess of neutrinos from low Galactic latitudes (see Fig.~\ref{fig:baikal_cascade_galactic}) \cite{Baikal_Galactic_median}.
This result adds up to the evidence for a Galactic neutrino flux which also builds up in the IceCube data \cite{Baikal_Galactic_median}.

\begin{figure}
  \centering
  \includegraphics[height=7.5cm]{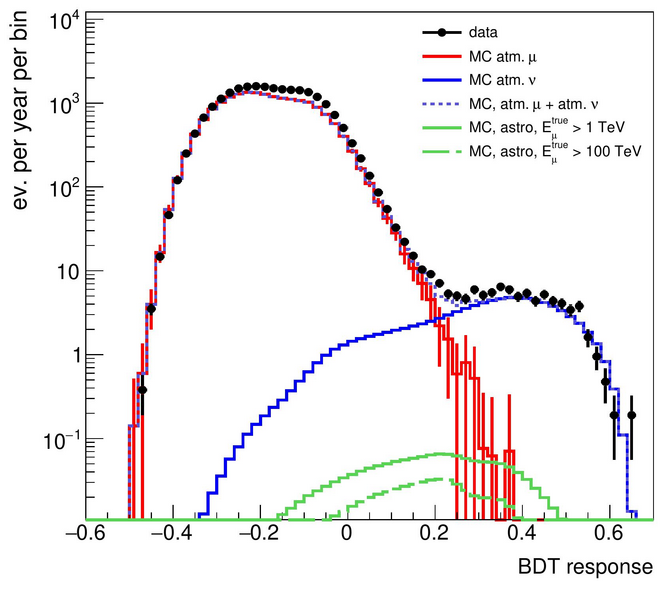}
  \includegraphics[height=7.5cm]{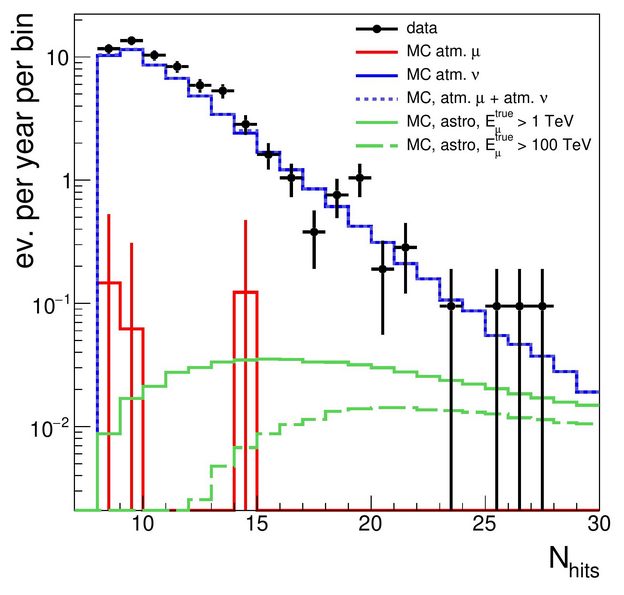}
  \caption{Results from a track analysis of the Baikal-GVD data collected in 2020--2021.
           Left: The distribution of the BDT classifier value for tracks reconstructed as upward-going. The MC expectations for atmospheric muons and atmospheric neutrinos are shown in red and blue, respectively.
The experimental data is shown by black points with statistical error bars.
           Right: The distribution of the number of hits after a cut on the BDT value (for neutrino candidate events).
  }
  \label{fig:track_analysis_results}
\end{figure}

The analysis of track-like events primarily focuses on upward-going events in order to suppress the large atmospheric muon background.
The current status of the track analysis is illustrated in Fig.~\ref{fig:track_analysis_results}, 
which shows results from a $\chi^2$-like track reconstruction applied to the 2020-2021 Baikal-GVD data.
Only events reconstructed as upward-going are considered.
As one can see, the distribution of the boosted decision tree (BDT) classifier, which is used to suppress the remaining atmospheric muon background (events mis-reconstructed as upward-going), shows a reasonable agreement between the Monte Carlo (MC) predictions and the experimental data.
After a cut on the BDT value, a high-purity set of neutrino candidate events is obtained.
The neutrino candidate event sample is characterized by a good data--MC agreement for all variables used in the training of the BDT classifier, e.g. for the number of hits which is shown in Fig.~\ref{fig:track_analysis_results}.
In the 2020-2021 data, 671 neutrino candidate events are selected in this analysis.
This event set is dominated by atmospheric neutrinos.
The analysis of the neutrino energy distribution, which would allow for isolation of the diffuse astrophysical neutrino flux, is in preparation.
The neutrino candidate events from all observing seasons are currently being used in various searches for point-like and extended neutrino sources, so far without any statistically significant detection.

\section{Conclusion}
\label{sect:conclusion}
Neutrino astronomy offers a unique view of the Universe.
This relatively young field of astronomy, which has been dominated by IceCube for many years, is now receiving a boost thanks to the ongoing construction of the KM3NeT telescopes in the Mediterranean Sea and the Baikal-GVD telescope in Lake Baikal.
These new telescopes complement IceCube in terms of field of view, while also providing an improved angular resolution.
The construction of Baikal-GVD has been steadily progressing in the past 8 years.
As of November 2024, Baikal-GVD consists of 4104 optical modules installed on 114 strings (compare to 594 OMs on 33 strings for KM3NeT-ARCA).
With a 0.6 km$^3$ effective volume (for cascades with E$>$100 TeV), Baikal-GVD is currently the largest neutrino telescope in the Northern Hemisphere.
The telescope has been collecting data in partial configurations since 2016.

The existence of a diffuse astrophysical neutrino flux, which was first revealed by IceCube, has now been confirmed by Baikal-GVD with a $>3\sigma$ significance \cite{Baikal_cascades}.
Baikal-GVD has also provided new evidence for a galactic neutrino flux \cite{Baikal_Galactic_median}, as well as further hints for neutrino flux from TXS 0506+056 and some other objects \cite{Baikal_cascades_MNRAS_2023,Baikal_TXS_cascade}.
The construction of Baikal-GVD is moving forward towards a 1~km$^3$ goal.

\section*{Acknowledgements}
This work is supported in the framework of the State project ``Science'' by the Ministry of Science and Higher Education of the Russian Federation under the contract 075-15-2024-541.


\begin{thebibliography}{9}
\section*{REFERENCES}

\bibitem{Markov1960} M.~A.~Markov,
in Proceedings of the 1960 Annual International Conference on High Energy Physics at Rochester,
Ed. by E.~C.~G. Sudarshan, J.~H.~Tinlot, and A.~C.~Melissinos (Univ. of Rochester, Rochester, 1960), p. 578.

\bibitem{IceCube} IceCube Collab. (M. G. Aartsen \emph{et~al.}), 
JINST {\bf 12}, P03012 (2017).

\bibitem{KM3NeT_LoI}
KM3NeT Collab. (S.~Adri\'an-Mart\'inez \emph{et~al.}),
J.\ Phys.\ G {\bf 43}, 084001 (2016).

\bibitem{ANTARES} ANTARES Collab. (M. Ageron \emph{et~al.}),
Nucl. Instrum. Methods A {\bf 656}, 11 (2011).

\bibitem{IceCube_diffuse}
IceCube Collab. (M. G. Aartsen \emph{et~al.}), 
Science {\bf 342}, 1242856 (2013).

\bibitem{IceCube_diffuse_2020}
IceCube Collab. (M. G. Aartsen \emph{et~al.}),
Phys.\ Rev.\ Lett. {\bf 125}, 121104 (2020).

\bibitem{IceCube_TXS0506} IceCube Collab.,
Science {\bf 361}, 147 (2018).

\bibitem{IceCube_TXS0506_flare} IceCube Collab.,
ApJL {\bf 863}, L30 (2018).

\bibitem{IceCube_NGC1068} IceCube Collab.,
Science {\bf 378}, 538 (2022).

\bibitem{IceCube_Galactic_diffuse} IceCube Collab.,
Science {\bf 380}, 1338 (2023).

\bibitem{Baikal_optical_water_properties}
A.~D. Avrorin \emph{et~al.} (Baikal-GVD Collab.), 
Nucl.\ Instrum.\ Meth.\ A {\bf 693}, 186 (2012).

\bibitem{Baikal_calibration} 
A.~D. Avrorin \emph{et~al.} (Baikal-GVD Collab.), 
PoS (ICRC2019) 0878, arXiv:1908.05458.

\bibitem{Baikal_positioning} 
A.~D. Avrorin \emph{et~al.} (Baikal-GVD Collab.), 
PoS (ICRC2019) 1012, arXiv:1908.05529.

\bibitem{Zaborov_Kleimenov_2024}
M. I. Kleimenov and D. N. Zaborov,
PEPAN Letters {\bf 21}, 646 (2024).

\bibitem{Baikal_cascades} 
V.~A.~Allakhverdyan \emph{et~al.} (Baikal-GVD Collab.), 
Phys.Rev.D {\bf 107}, 042005 (2023).

\bibitem{Baikal_cascades_MNRAS_2023} 
V.~A.~Allakhverdyan \emph{et~al.},
MNRAS {\bf 526}, 942 (2023).

\bibitem{Baikal_TXS_cascade}
V.~A.~Allakhverdyan \emph{et~al.},
MNRAS {\bf 527}, 8784 (2024).

\bibitem{Baikal_Galactic_median}
V.~A.~Allakhverdyan \emph{et~al.}, 
submitted to PRL, arXiv:2411.05608.

\end{thebibliography}
\end{document}